\documentclass[prl,twocolumn,showpacs,amsmath,amssymb,aps]{revtex4}

\usepackage{graphicx}
\usepackage{dcolumn}
\usepackage{bm}

\newcommand\mean[1]{{\langle#1\rangle}}
\newcommand{\Ocal}{{\cal O}}

\newcommand{\Pcal}{{\cal P}}
\newcommand{\rhotdt}{\rho_{t+\Delta t}}
\newcommand{\dwt}{\Delta W_t}

\begin{document}





\title{Numerical schemes for continuum models of reaction-diffusion systems\\
subject to internal noise \vspace{-0.2cm}}



\author{Esteban Moro} \email{emoro@math.uc3m.es}
\affiliation{%
Grupo Interdisciplinar de Sistemas Complejos (GISC) and  Departamento de Matem\'aticas, Universidad Carlos III de
  Madrid, Avda.\ de la Universidad 30, E-28911 Legan\'es, Spain
}%

\date{\today}

\begin{abstract}
We present new numerical schemes to integrate stochastic partial
differential equations which describe the spatio-temporal dynamics
of reaction-diffusion (RD) problems under the effect of internal
fluctuations. The schemes conserve the nonnegativity of the
solutions and incorporate the Poissonian nature of internal
fluctuations at small densities, their performance being limited
by the level of approximation of density fluctuations at small
scales. We apply the new schemes to two different aspects of the
Reggeon model namely, the study of its non-equilibrium phase
transition and the dynamics of fluctuating pulled fronts. In the
latter case, our approach allows to reproduce quantitatively for
the first time microscopic properties within the continuum model.
\end{abstract}



\maketitle

Continuum representations of the dynamics of spatially-extended
systems subject to fluctuations is a very active area of research
in statistical mechanics and nonlinear dynamics
\cite{sancho,barabasi,munoz,hohenberg,gardiner}. This is because
they are frequently more tractable than discrete models, they can
be put forward using simple symmetry arguments and applying
conservation laws, and therefore they provide minimal
representations of the observed phenomena. Important instances are
Langevin equations for the relaxational dynamics of equilibrium
models \cite{hohenberg}, growth interface phenomena
\cite{barabasi} or coarse-grained descriptions of microscopic RD
problems \cite{gardiner,hinrichsen}. Despite their apparent
simplicity, most of these models can not be solved analytically
and one has to resort to approximate analytical techniques, or to
numerical integration of the stochastic time-dependent set of
equations using well established algorithms \cite{kloeden}. In the
important instance of RD systems subject to internal fluctuations
the configurations are given by a {\em non-negative} density field
$\rho(x,t)$ subject to fluctuations of typical strength
$\sqrt{\rho(x,t)}$ which accounts for the {\em Poissonian
fluctuations} of the number of particles at $x$ \cite{gardiner}.
Unfortunately, standard algorithms fail to guarantee both the
essential non-negativity of $\rho(x,t)$ and the Poissonian
character of its fluctuations. Our purpose in this paper is to
propose efficient numerical algorithms to overcome these problems
which will allow us to prove the importance of internal
fluctuations and to check the relevance of their correct
description at different scales. 

In this paper we concentrate in the so called Reggeon model, which
in one dimension is given by \cite{hinrichsen}
\begin{equation}\label{eq1}
\frac{\partial \rho}{\partial t}=D\frac{\partial^2 \rho}{\partial
x^2} + \rho - \rho^2 + \sqrt{\sigma \rho}\ \eta(x,t),
\end{equation}
where $\eta(x,t)$ is a Gaussian white noise. The Reggeon model can
be obtained under some approximations from the microscopic Master
equations of RD microscopic models using well-known techniques
\cite{gardiner,doi}. Heuristically, Eq.\ (\ref{eq1}) can also be
considered as the simplest dynamical equation for a coarse-grained
density field with $\sigma = 1/N$, $N$ being the mean-field number
of particles per site. The Reggeon model provides a minimal
representation of the Directed Percolation (DP) universality
class, which is currently regarded as paradigm of non-equilibrium
systems with absorbing states \cite{hinrichsen}: if $\bar \rho(t)$
is the mean density spatial average, there exists a critical value
of $\sigma$ for which (\ref{eq1}) undergoes a transition between
an active phase $\lim_{t\to\infty} \bar \rho(t) \neq 0$ and an
{\em absorbing phase} for which $\lim_{t\to\infty} \bar \rho(t) =
0$.

In addition, when $\sigma = 0$ Eq.\ (\ref{eq1}) becomes the so
called Fisher-Kolmogorov-Petrovsky-Piscounov (FKPP) equation
\cite{Fisher}, which displays {\em pulled fronts} in which the
active phase invades the absorbing state
\cite{saarloos,brunet,pechenik,doering}. Simulations of
microscopic particle models \cite{moro,brunet} have shown that the
dynamics of pulled fronts are extremely sensitive to microscopic
fluctuations at $\rho \simeq 1/N$, leading to strong corrections
in the front properties when compared with those of the FKPP
equation. Since Eq.\ (\ref{eq1}) is usually held as a continuum
description of some particle models at the mesoscopic level (i.e.\
when $\rho \gg 1/N$) one might doubt that the Reggeon model
describes correctly the behavior of pulled fronts subject to
internal fluctuations. The efficiency and accuracy of the
numerical schemes proposed here will allow us to show that Eq.\
(\ref{eq1}) indeed incorporates the ingredients to explain (even
quantitatively) the phenomena observed in particle models, thus
providing also a minimal representation of pulled fronts subject
to internal fluctuations.

To simplify the discussion, let us consider the simplest possible
case for the dynamics of a density subject to internal
fluctuations:
\begin{equation}\label{eq0}
\frac{d\rho}{dt} = a\rho + \sqrt{\sigma \rho} \: \eta(t).
\end{equation}
Typical explicit or implicit methods based on stochastic Taylor
approximations of (\ref{eq0}) immediately run into problems, since
they do not conserve the nonnegativity of $\rho(t)$. For example,
the Euler approximation is \cite{kloeden}
\begin{equation}\label{euler}
\rhotdt=\rho_t + a \rho_t \Delta t + \sqrt{\sigma \rho_t} \dwt,
\end{equation}
where $\dwt$ are random Gaussian numbers with zero mean and
$\Delta t$ variance. Thus, there is a finite probability that
$\rhotdt$ becomes negative, and the numerical integration comes to
a halt. In order to overcome this problem, Dickman proposed an
interesting solution based on the Euler scheme (\ref{euler}) and
the discretization of the possible values of $\rho_t$ as multiples
of $\rho^* = \Ocal(\sigma\Delta t)$ \cite{dickman}. Despite its
success in reproducing the universality class exponents of DP
using (\ref{eq1}) and its application to other situations
\cite{dickman}, Dickman's algorithm is not really a numerical
integration of a continuum model. Moreover, no general study of
its convergence and applicability for other situations has been
done yet. A more technical solution was proposed by Schurz and
coworkers \cite{schurz1} using Balanced Implicit Methods (BIM), in
which implicit Euler methods are used to impose the nonnegativity
of the solution. In the case of Eq.\ (\ref{eq0}) the BIM scheme
reads \cite{note1}
\begin{equation}\label{bim}
\rhotdt = \frac{\rho_t + \Delta t \rho_t + \sqrt{\sigma \rho_t}\:
(\dwt + |\dwt|)}{1+\sqrt{\sigma/\rho_t}\: |\dwt|},
\end{equation}
which explicitly implements the constraint $\rhotdt \geq 0$, and
reduces to the Euler algorithm (\ref{euler}) up to order
$\Ocal(\Delta t)$ \cite{schurz1,moronew}. The BIM scheme is known
to have the same order of convergence as the Euler algorithm,
namely, the error is $\Ocal(\sqrt{\Delta t})$ for approximations
of individual trajectories and $\Ocal(\Delta t)$ for moments of
$\rho(t)$ \cite{kloeden,schurz1}.

Another approach was taken by Pechenik and Levine \cite{pechenik}
employing the exact conditional probability density (CDF)
$\Pcal(\rhotdt|\rho_t)$ for the stochastic process satisfying
(\ref{eq0}), which has been known for some time in economy as the
Cox-Ingersoll-Ross process \cite{cox}. The CDF can be expressed in
terms of modified Bessel functions and, although it can be sampled
numerically using rejection or transformation methods
\cite{pechenik}, it is computationally expensive. Here we propose
a more efficient procedure, which is based on the following: if we
define $r_d(t) = \sum_{i=1}^d x_i^2(t)$, where $x_i(t)$ satisfies
$d x_i/dt = a x_i/2 + (\sigma/4)^{1/2}\eta_i(t)$ with $\eta_i(t)$
independent white noises, then $d r_d/dt = d\sigma/4+ar_d+(\sigma
r_d)^{1/2} \eta(t)$ which coincides with (\ref{eq0}) in the limit
$d\to0$. Since the equation for each $x_i(t)$ is linear, $r_d(t)$
is the sum of squares of Gaussian random numbers with non-zero
mean. Thus its probability distribution is related to the $\chi^2$
distribution with $d$ degrees of freedom \cite{libro}.
Specifically, we find that $\rhotdt = r_0(t+\Delta t) =
\chi'^2_0(\lambda)/(2k)$ where $k=2a/[\sigma (e^{a\Delta t}-1)]$,
$\lambda = k e^{a\Delta t}\rho_t$ and $\chi'^2_0(\lambda)$ is a
random number with a noncentral $\chi^2$ distribution with zero
degrees of freedom and noncentrality parameter $\lambda$ whose
cumulative distribution function is given by
\cite{libro,moronew,munoznew}
\begin{equation}\label{cdf}
\Pcal[\chi^2_0(\lambda) \leq x] = \sum_{j=1}^\infty
\frac{(\lambda/2)^j e^{-\lambda/2}}{j!} \Pcal[\chi^2_{2 j}\leq x]+
e^{-\lambda/2} \Theta(x),
\end{equation}
where $\chi^2_{2j}$ is a $\chi^2$ random number with $2j$ degrees
of freedom and $\Theta(x)$ is the step function. Equation
(\ref{cdf}) is important for two reasons: {\em (i)} it shows that
there is a finite probability $ \Pcal(\rhotdt = 0) =
e^{-\lambda/2}$ for getting into the absorbing state, and more
importantly {\em (ii)} it reveals that the probability
distribution of $\chi'^2_0(\lambda)$ is a linear combination of
$\chi^2$ probability distributions with Poisson weights. This fact
can be exploited to generate $\rhotdt$ efficiently: if we choose
$K$ from a Poisson distribution with mean $\lambda/2$, then
\begin{equation}\label{rhodett}
\rhotdt = \frac{1}{2k} \left\{\begin{array}{ll}0 & \mathrm{if}\ K
= 0,
\\ \sum_{i=1}^{2K} z_i^2 & \mathrm{if}\ K \neq 0, \end{array}\right.
\end{equation}
where $z_i$ are independent Gaussian random numbers with zero mean
and unit variance.

\begin{figure}
\begin{center}
\includegraphics[width=3.0in,clip=]{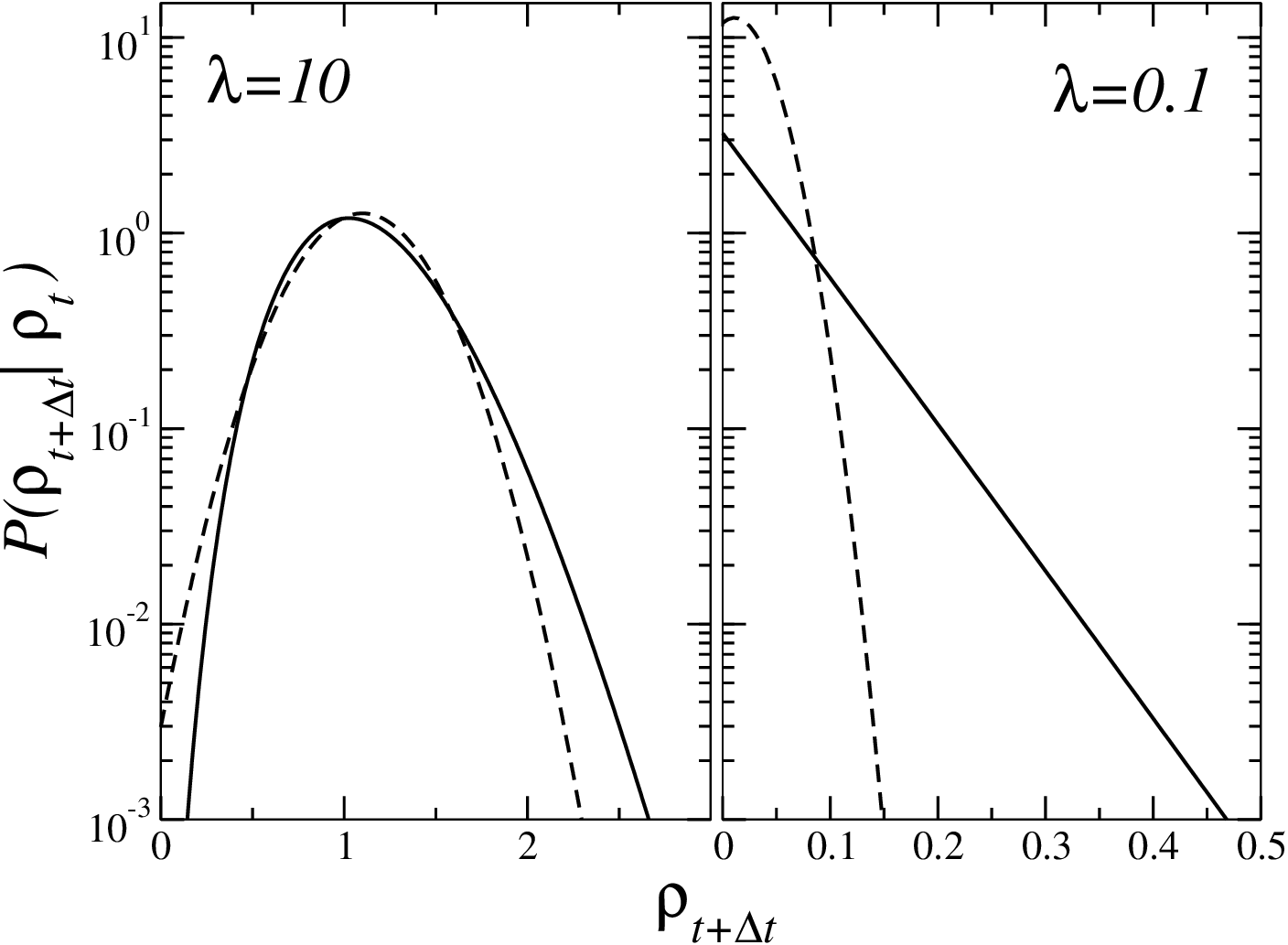}
\caption{\label{fig-pdf} Conditional probability density as a
function of $\rhotdt$ for Eq.\ (\ref{eq0}) with $\Delta t = 0.1$,
$\rho_t = 1$ (left) and $\rho_t = 10^{-2}$ (right). Solid lines
are the exact solution from (\ref{cdf}), while dashed lines are
the approximations obtained using the Euler scheme (\ref{euler}).
\vspace{-0.5cm}}
\end{center}
\end{figure}

Another interesting feature of the exact CDF for (\ref{eq0}) is
the fact that it converges asymptotically towards the Euler
approximation (\ref{euler}) when $\lambda \simeq \rho_t/(\sigma
\Delta t) \gg 1$ \cite{libro,moronew}. However, for small
$\lambda$, the Euler approximation underestimates the large
fluctuations present in the exact solution of (\ref{eq0}). This
effect, which can be seen in Fig.\ \ref{fig-pdf}, is related to
the fact that the Gaussian approximation (\ref{euler}) of a
Poisson random number (\ref{eq0}) is only valid when the mean
value is large enough \cite{moronew}. The failure of
approximations like (\ref{euler}) or (\ref{bim}) to reproduce
large density fluctuations at small values of $\lambda$ introduces
an effective microscopic cutoff $\rho^* = \Ocal(\sigma \Delta t)$
in the numerical simulations below which these approximations
break down.

Although the scheme (\ref{bim}) can be easily generalized to
integrate equations like (\ref{eq1}), this is not the case for the
exact sampling of the CDF for (\ref{eq0}). Thus, a {\em
splitting-step} strategy for integrating Eq.\ (\ref{eq1}) was
proposed in \cite{pechenik}, where the time interval $\Delta t$ is
split into two steps: (i) given $\rho_t$, we use (\ref{rhodett})
to integrate (\ref{eq0}) and get an intermediate value ${\tilde
\rho}_{t+\Delta t}$; (ii) we take $\tilde \rho_{t+\Delta t}$ as
the initial condition for $
\partial \rho/\partial t = \partial^2 \rho/\partial
x^2 - \rho^2$, producing $\rhotdt$ with the aid of any
deterministic numerical algorithm. It can be proved that this
splitting step method (SSM) converges towards the solutions of
(\ref{eq1}), its order of convergence being $\Ocal(\Delta t)$ both
for realizations and for moments of $\rho(t)$ \cite{moronew}. This
means that the splitting-step method provides better
approximations than those based on Euler methods (like the Dickman
and BIM algorithms) for any realization of the noise. This has
significant consequences when characterizing the critical point,
as will be shown below. In the following we apply the two methods
proposed here [BIM and the SSM using (\ref{rhodett})] and the
Dickman algorithm to the two problems for which Eq.\ (\ref{eq1})
is archetypal \footnote{In our simulations we have used $D=1$ and
a spatial discretization with $\Delta x=1$ in a one dimensional
lattice with $L$ nodes.}.

\begin{figure}
\begin{center}
\includegraphics[width = 3.0in,clip=]{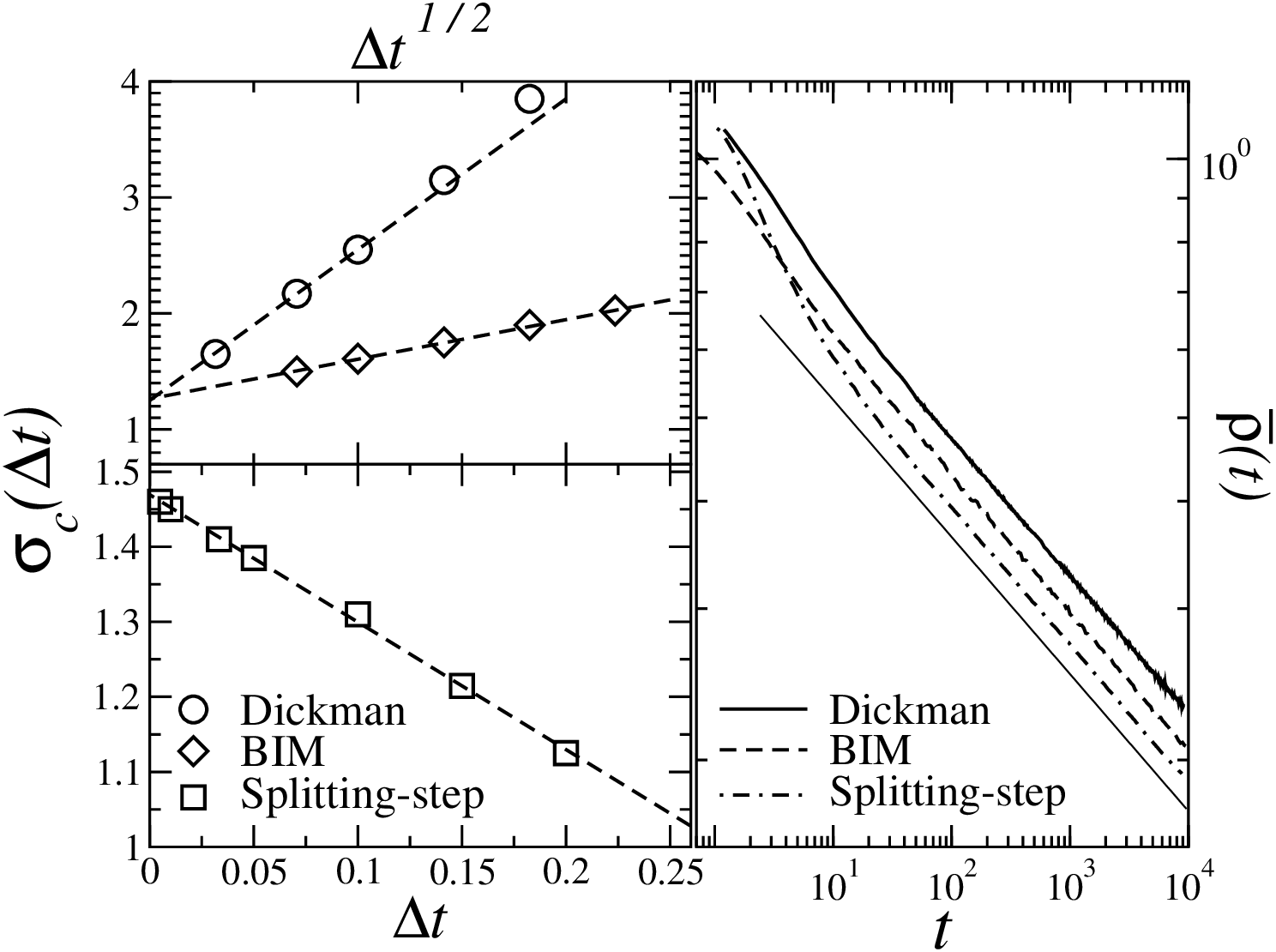}
\caption{\label{fig-tc} {\em Left:} convergence analysis for the
different algorithms. Points are the critical value of $\sigma$ as
a function $\Delta t$ (below) and of $\Delta t^{1/2}$ (up) while
dashed lines are linear fits to the data. System size is $L=400$.
{\em Right:} time dependence of the mean density $\bar \rho (t)$
at the critical point $\sigma = \sigma_c(\Delta t)$ with $\Delta
t=10^{-2}$ obtained using the different algorithms (lines are
shifted vertically for clarity). Thin line is the power law $\bar
\rho(t) \sim t^{-\delta}$ with $\delta = 0.1595$. System size is
$L=1000$. \vspace{-0.5cm}}
\end{center}
\end{figure}

{\em Study of the DP phase transition.} To test the proposed
algorithms, we study the well known non-equilibrium phase
transition that Reggeon model displays for moderate values of
$\sigma$ \cite{dickman}. At the critical point, the mean average
density $\bar \rho(t) \equiv \frac{1}{L} \sum_{x}
\mean{\rho(x,t)}$ decays like a power law $\bar \rho \sim
t^{-\delta}$ with $\delta\simeq 0.1595$ \cite{hinrichsen}. As in
\cite{dickman}, we identify the critical point as the value of
$\sigma$ for which we observe such a power law decay in $\bar
\rho(t)$. Results for the different algorithms are shown in Fig.\
\ref{fig-tc}, where we report the value of $\sigma_c$ as a
function of the time step $\Delta t$. As expected, the order of
convergence of the Dickman and BIM methods is $\Ocal(\sqrt{\Delta
t})$, while the SSM has $\Ocal(\Delta t)$ order of convergence.
The improvement in the order of convergence comes with a price:
the computer time needed for our numerical simulations at the
critical point (see table \ref{table1}) indicate that methods
based on Euler approximations, despite having an effective
microscopic cutoff at $\rho^* = \Ocal(\sigma \Delta t)$, are
faster than the SSM, and thus could provide better strategies for
integrating numerically equations for RD models close to the
critical point, where only accurate approximations of {\em large
length and time scales} are needed.

\begin{table}
\caption{\label{table1} Comparison of CPU time at $\sigma =
\sigma_c(\Delta t)$ with $\Delta t= 10^{-2}$ and $L=400$ for the
different algorithms in Fig.\ \ref{fig-tc}, normalized to that of
Dickman's algorithm.}
\begin{ruledtabular}
\begin{tabular}{lcc}
Method&$\sigma_c$& $T_{run}$\\
\hline
Dickman & 2.55 & 1\\
BIM & 1.61 & 1.2\\
Splitting-Step & 1.45 & 7.6
\end{tabular}
\end{ruledtabular}
\end{table}

{\em Dynamics of fluctuating pulled fronts.} When $\sigma = 0$,
equation (\ref{eq1}) displays a wave-like solution (front) which
travels with velocity $v_0 = 2\sqrt{D}$ (provided sharp enough
initial conditions are given) \cite{Fisher,saarloos}. The dynamics
of this {\em pulled front} is severely affected when microscopic
fluctuations close to the absorbing state $\rho = 0$ are
considered. Specifically, it has been observed in particle models
whose mean field limit is given by the FKPP equation, that the
front speed is universally modified as \cite{brunet,saarloos,moro}
\begin{equation}\label{bd1}
v_N \equiv \lim_{t\to \infty} \frac{\mean{x_f(t)}}{t} \simeq v_0 -
v_0 \frac{C}{\ln^2 N},
\end{equation}
where $x_f(t)$ is the instantaneous position of the front, $C$ is
a positive constant and $N$ is the number of particles per site
\cite{moro}. Moreover, the pulled front diffuses with diffusion
constant
\begin{equation}\label{bd2}
D_{f,N} \equiv \lim_{t\to \infty} \frac{\mean{(x_f(t)- v_N
t)^2}}{2t} \simeq \frac{C'}{\ln^3 N},
\end{equation}
where $C'$ is a positive constant. Whereas the velocity correction
can be easily understood because microscopic fluctuations at $\rho
\simeq N^{-1}$ provide an effective cutoff in the dynamics
\cite{brunet}, the diffusion coefficient seems to depend on the
existence of relatively large fluctuations in the density at $\rho
\simeq N^{-1}$ and on their slow relaxation by the pulled front
dynamics \cite{moro}.

As mentioned in the introduction, one might doubt that the large
microscopic fluctuations at $\rho = N^{-1}$ observed in particle
models are correctly reproduced by such type of equation like
(\ref{eq1}). Note, however, that the relationship between particle
models and the Reggeon field model is deeper than at the
coarse-grained level. Specifically, in \cite{doering} it was shown
that there is an {\em exact duality transformation} between the $A
\leftrightarrow A+A$ microscopic particle model and the so called
stochastic FKPP equation, which is similar to the Reggeon model
but with a $\sqrt{\sigma \rho(1-\rho)}\:\eta(x,t)$ noise term. For
$\sigma \ll 1$, the noise is only relevant at very small values of
$\rho$ where $\sqrt{\sigma \rho(1-\rho)} \simeq \sqrt{\sigma
\rho}$ and thus, both the Reggeon model and the stochastic FKPP
should provide similar results.

\begin{figure}
\begin{center}
\includegraphics[width=3.0in,clip=]{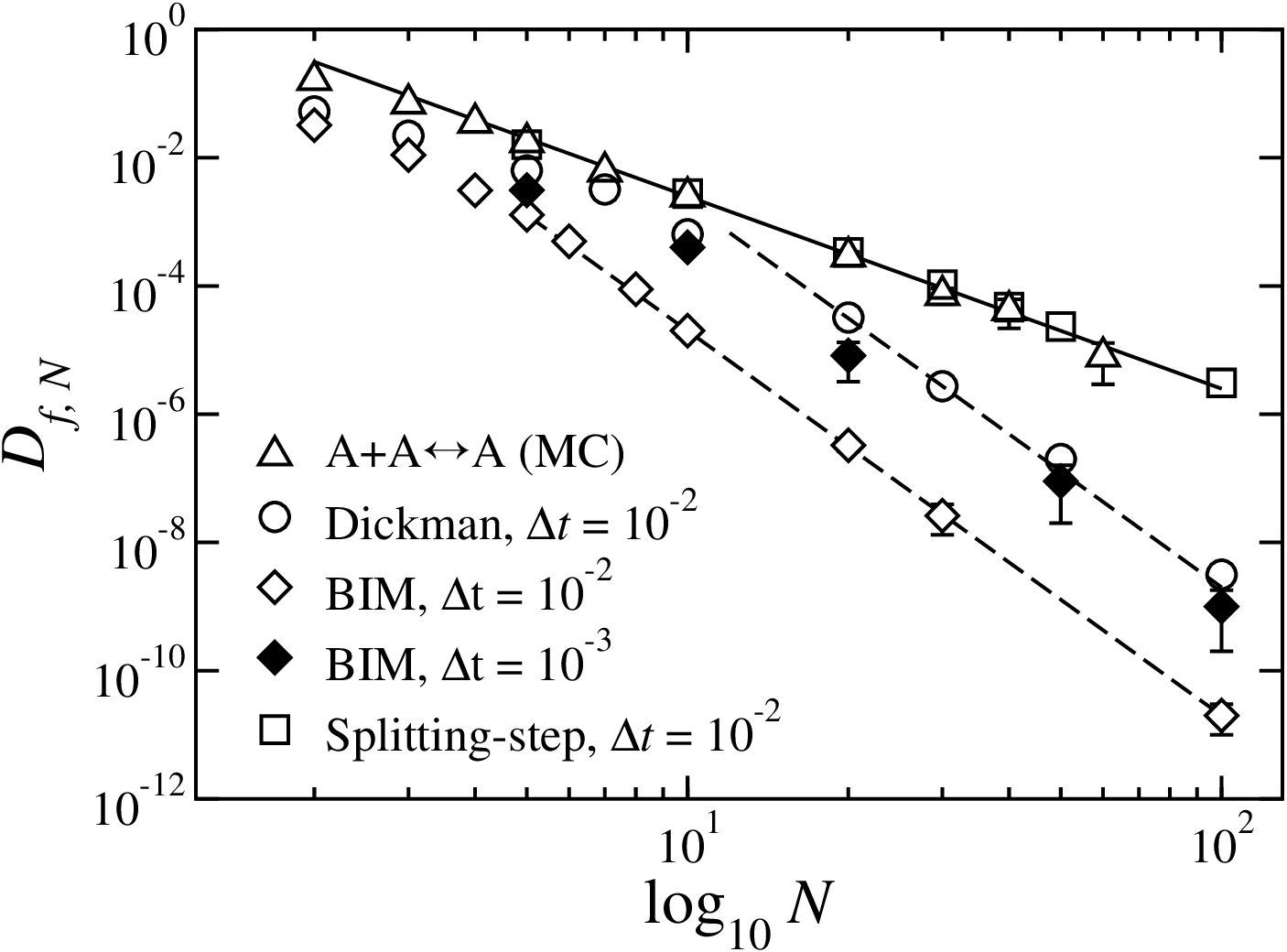}
\caption{\label{fig-dif} Front diffusion coefficient as a function
of $\sigma$ for the different algorithms and different $\Delta t$,
compared with hybrid MC simulations of the microscopic model
$A\leftrightarrow A+A$ \cite{moro}. Solid (dashed) line is the
$\log^{-3} N$ ($\log^{-6} N$) power law. \vspace{-0.5cm}}
\end{center}
\end{figure}

Our results for the front diffusion coefficient, obtained by
numerical integration of Eq.\ (\ref{eq1}) are reported in Fig.\
\ref{fig-dif} together with those of hybrid Monte-Carlo results
for the $A \leftrightarrow A+A$ particle model \cite{moro}. As we
can see, for a given time step $\Delta t$, the SSM reproduces the
$\ln^{-3} N$ results for particle models (\ref{bd2}) thus
confirming the duality relationship between the particle model and
the continuum equation even at the quantitative level. However,
the other algorithms are more consistent with a $\ln^{-6} N$
scaling which, interestingly, can be obtained through standard
perturbation techniques based on Gaussian approximations for the
fluctuations of the front position \cite{saarloos}. The reason for
this difference among the various schemes is related to the fact
that both the Dickman and BIM algorithms are based on Gaussian
approximations for the density fluctuations which are
underestimated for $\rho < \rho^*=\Ocal(\sigma \Delta t)$, while
only the SSM reproduces exactly the large density Poissonian
fluctuations (also observed in particle models) when $\rho$ is
small. This does not mean that the Dickman and BIM algorithms do
not converge in this case: specifically, if we take $\Delta t \to
0$ we observe that the value of the diffusion coefficient
approaches that of the hybrid MC simulations for the $A
\leftrightarrow A + A$ (see Fig.\ \ref{fig-dif}). Thus the
applicability of the Dickman and BIM algorithms is limited in this
case since they fail to reproduce fluctuations at {\em small
density and time scales}.

In summary, we have presented new strategies for integrating
stochastic (partial) differential equations for models of RD
subject to internal fluctuations. While all of them preserve the
nonnegativity of the solution, algorithms based on Gaussian
approximations introduce a microscopic cutoff below which density
fluctuations are not correctly accounted for. This is not
important when the system properties are dominated by the dynamics
of large length and time scales (as in critical behavior), and
thus, schemes based on Euler approximation suffice to integrate
numerically equations like (\ref{eq1}). However, when the observed
phenomena are sensitive to microscopic fluctuations, only
algorithms which take into account the exact sampling of density
fluctuations at small scales are computationally efficient.
Moreover, our results validate continuum models like (\ref{eq1})
to study the dynamics of fluctuating pulled fronts and corroborate
the importance of Poissonian large fluctuations of the density at
small scales. We hope that our results will be used in future for
the analytical understanding of pulled front dynamics
\cite{saarloos,moro}.

Finally, we mention that the methods presented here (the BIM and
SSM) can be easily extended to other situations in which the
relevant degrees of freedom are non-negative
\cite{munoznew,moronew}, like the study of density fluctutions in
more general RD problems \cite{gardiner}, the understanding of
critical phenomena of systems subject to external/multiplicative
noise (e.g. with $\rho\: \eta(x,t)$ noises) \cite{sancho,munoz},
or the nonlinear modelling of the behavior of interest rates in
economy \cite{cox,schurz1}.

  We are grateful to E.\ Brunet, R.\ Cuerno, C.\ Doering,
  H.\ Schurz, and P.\ Smereka for comments and
  discussions. Financial support is acknowledged from
  the Ministerio de Ciencia y Tecnolog\'\i a (Spain).


\vspace{-0.3cm}


\begin{thebibliography}{99}
\bibitem{hohenberg} M.\ C.\ Cross and P.\ C.\ Hohenberg, Rev.\ Mod.\ Phys.\ {\bf 65}, 851 (1993).
\bibitem{sancho} J. Garc\'{\i}a-Ojalvo and J. Sancho, {\em Noise
in spatially Extended Systems} (Springer-Verlag, New York, 1999).
\bibitem{munoz} M.\ A.\ Mu\~noz, in {\em Advances in Condensed
Matter and Statistical Mechanics}, E.\ Korutcheva and R.\ Cuerno
Eds. (Nova Sience, New York, 2004).
\bibitem{barabasi} A.-L. Barab\'asi and H. E. Stanley, {\em Fractal
concepts in surface growth} (Cambridge Univ. Press, Cambridge,
1995); J. Krug, Adv. Phys. {\bf 46}, 129 (1997).
\bibitem{gardiner} C. W. Gardiner, {\em Handbook of Stochastic Methods},
(Springer, Berlin, 1996).
\bibitem{hinrichsen} H.\ Hinrichsen, Adv.\ Phys.\ {\bf 49}, 815 (2000).
\bibitem{kloeden} P.\ E.\ Kloeden, E.\ Platen, {\em Numerical Solution of Stochastic Differential Equations} (Springer-Verlag, 1992).
\bibitem{doi} M. Doi, J. Phys. A {\bf 9}, 1479 (1976); L. Peliti,
J. Phys. (Paris) {\bf 46}, 1469 (1985); D. C. Mattis and M. L.
Glasser, Rev. Mod. Phys. {\bf 70}, 979 (1998).
\bibitem{Fisher} R. A. Fisher, Ann. Eugenics {\bf VII}, 355 (1936);
A. Kolmogorov, I. Petrosvky, and N. Piscounov, Moscow Univ. Bull.
Math. A {\bf 1}, 1 (1937);
\bibitem{saarloos} W. van Saarloos, Phys.\ Rep.\ {\bf 386} 29 (2003); D.\ Panja,
{\em ibid}, {\bf 393}, 87 (2004).
\bibitem{brunet} E.\ Brunet and B.\ Derrida, Phys.\ Rev.\ E {\bf 56}, 2597 (1997); J.\ Stat.\ Phys.\ {\bf 103}, 269 (2001).
\bibitem{pechenik} L.\ Pechenik and H.\ Levine, Phys.\ Rev.\ E {\bf 59}, 3893 (1999).
\bibitem{doering} C.\ R\ Doering, C.\ Mueller, and P.\ Smereka, Physica A {\bf 325}, 243 (2003).
\bibitem{moro} E.\ Moro, Phys.\ Rev.\ Lett.\ {\bf 87} 238303 (2001);
Phys.\ Rev.\ E {\bf 69}, 060101 (2004)
\bibitem{munoznew} While writing this paper we became aware of a
recent preprint, I.\ Dornic, H.\ Chat\'e, M.\ A.\ Mu\~noz,
cond-mat/0404105, in which a splitting-step like the one in
\cite{pechenik} and a similar procedure to compute $\rhotdt$ using
(\ref{cdf}) is given.
\bibitem{dickman} R.\ Dickman, Phys.\ Rev.\ E {\bf 50}, 4409
(1994); C. L\'opez and M.\ A.\ Mu\~noz, {\em ibid} {\bf 56}, 4864
(1997).
\bibitem{schurz1} G.\ N.\ Milshtein, E.\ Platen, and H.\ Schurz, SIAM J.\ Numer.\ Anal.\ {\bf 35}, 1010 (1998);
H.\ Schurz, Dyn.\ Syst.\ Appl.\ {\bf 5}, 323 (1996).
\bibitem{note1} Convergence of scheme (\ref{bim}) requires a cut-off at $\rho
\simeq \varepsilon$, where $\varepsilon$ is chosen small enough
(see \cite{schurz1}).
\bibitem{moronew} E.\ Moro, in preparation.
\bibitem{cox} J.\ Cox, E.\ Ingersoll, and S.\ A.\ Ross,
Econometrics {\bf 53}, 385 (1985).
\bibitem{libro} N.\ L.\ Johnson, S.\ Kotz and N.\ Balakrishnan,
{\em Continuous univariate distributions} (Vol. II) (John Wiley \&
Sons, New York, 1994); A.\ F.\ Siegel, Biometrika {\bf 66}, 381
(1979).
\end{thebibliography}
\end{document}